\documentclass[12pt]{iopart}
\usepackage{iopams}
\usepackage{epsfig}
\begin{document}

\title[Competition of tolerant strategies in the spatial public goods game]{Competition of tolerant strategies in the spatial public goods game}

\author{Attila Szolnoki$^1$ and Matja{\v z} Perc$^{2,3}$}
\address{$^1$Institute of Technical Physics and Materials Science, Centre for Energy Research, Hungarian Academy of Sciences, P.O. Box 49, H-1525 Budapest, Hungary\\
$^2$Faculty of Natural Sciences and Mathematics, University of Maribor, Koro{\v s}ka cesta 160, SI-2000 Maribor, Slovenia\\
$^3$CAMTP -- Center for Applied Mathematics and Theoretical Physics, University of Maribor, Krekova 2, SI-2000 Maribor, Slovenia}
\ead{szolnoki@mfa.kfki.hu, matjaz.perc@uni-mb.si}

\begin{abstract}
Tolerance implies enduring trying circumstances with a fair and objective attitude. To determine whether evolutionary advantages might be stemming from diverse levels of tolerance in a population, we study a spatial public goods game, where in addition to cooperators, defectors, and loners, tolerant players are also present. Depending on the number of defectors within a group, a tolerant player can either cooperate in or abstain from a particular instance of the game. We show that the diversity of tolerance can give rise to synergistic effects, wherein players with a different threshold in terms of the tolerated number of defectors in a group compete most effectively against defection and default abstinence. Such synergistic associations can stabilise states of full cooperation where otherwise defection would dominate. We observe complex pattern formation that gives rise to an intricate phase diagram, where invisible yet stable strategy alliances require outmost care lest they are overlooked. Our results highlight the delicate importance of diversity and tolerance for the provisioning of public goods, and they reveal fascinating subtleties of the spatiotemporal dynamics that is due to the competition of subsystem solutions in structured populations.
\end{abstract}


\maketitle

\section{Introduction}
The stability of modern human societies relies on public goods that future generations have, for many decades past, increasingly often taken for granted \cite{ostrom_90}. This is likely to have dire consequences, because we fail to grasp the long-term consequences of unsustainable exploitation of natural resources and the unchecked growth of consumerism as the dominating social and economic order \cite{deaton_80}. Tolerance, as the ability or willingness to steadfastly endure something, in particular a trying circumstance such as the existence of opinions or behaviour that one does not necessarily agree with, might be an important part of the equation to take us back on track. Although according to Darwin's \textit{The Origin of Species} natural selection favours the fittest and the most successful individuals, which in turn implies an innate selfishness that greatly challenges the concept of cooperation, tolerance and social norms in human societies may just be the missing ingredient for cooperative behaviour to prevail \cite{smith_10, nowak_11, chen_xj_pre09b, jiang_ll_pre10, rand_tcs13, helbing_pcbi10, xu_b_srep15, capraro_prsb15, biziou_jdm15, pavlogiannis_srep15}. In the realm of these considerations, evolutionary game theory waits ready as the theoretical foundation that is fit for addressing such challenges in a concise and relevant manner \cite{sigmund_93, weibull_95, hofbauer_98, nowak_06, sigmund_10}. The public goods game \cite{perc_jrsi13}, in particular, has been established for addressing situations that, if left unattended, may evolve towards the tragedy of the commons \cite{hardin_g_s68}.

The crux of the underlying social dilemma lies in the fact that cooperation foresees an altruistic act that is costly to perform but benefits another. Although mutual cooperation would be optimal for all involved, a higher payoff, at least in the short term, is possible by exploiting the cooperative efforts of others. An often explored remedy is to punish defectors who do not contribute to the public good \cite{fehr_aer00, gardner_a_an04, henrich_s06b, sigmund_tee07, raihani_tee12, rand_nc11, powers_jtb12, shimao_pone13, hauser_jtb14}, or to reward cooperators who do contribute to the public good \cite{dreber_n08, rand_s09, hilbe_prsb10, hauert_jtb10, szolnoki_epl10, szolnoki_njp12, szolnoki_prsb15,gao_l_srep15}. However, the problem with both actions is that they are costly. Punishment entails bearing a cost for somebody else to incur a cost, while rewarding entails bearing a cost for somebody else to incur a benefit. As such, the basic dilemma is simply transferred to another level, hence the term second-order free-riding \cite{fehr_n04, fowler_n05b}, which describes all those who benefit from either reward or punishment but do not contribute towards covering the related costs. Another option is to simply abstain from the game if the outlook is not sufficiently rosy, and to settle for a small, but secure payoff instead \cite{hauert_s02, szabo_prl02}. Previous research has shown, however, that defectors, cooperators, and loners become entailed in a closed loop of dominance \cite{kerr_n02, kirkup_n04, reichenbach_pre06, arenas_jtb11, wang_wx_pre11, szolnoki_jrsif14, groselj_pre15}, which maintains a Red Queen existence of cooperative behaviour that on average is no better than if everybody would abstain \cite{hauert_jtb02, semmann_n03}.

Here we consider diverse levels of tolerance in the public goods game as an alternative to the aforementioned strategies. Indeed, instead of simply abandoning a less prospective public goods enterprise by abstaining, it may make much more sense to tolerate the situation and to endure against an emerging negative trend. Thus, in addition to cooperators, defectors, and loners, we also consider tolerant players. They are those who monitor the strategies of their neighbors and behave accordingly. In particular, if the number of defectors in the group exceeds a certain threshold tolerant players behave as loners, while otherwise they cooperate. Such behavior is similar in spirit to conditional cooperation \cite{taylor_76, boyd_jtb88, szolnoki_pre12, nakamaru_pone14}, but in the present case we do not neglect the fact that monitoring others to know the actual number of defectors in the group requires an additional effort, and thus comes with an additional cost of inspection. Importantly, instead of the application of a single level of tolerance \cite{szolnoki_pre15}, the question is whether diverse tolerance strategies offer evolutionary advantages over default abstinence, and whether there is synergy or competition among them? As we will show, advantages to tolerance and synergies do exist, such that players with a different threshold in terms of the tolerated number of defectors in a group provide an optimal response to the public goods dilemma. These solutions are made possible by the spontaneous emergence of complex spatial patterns that require a sufficiently large system size to emerge from random initial conditions. In fact, we will show that some globally stable subsystem solutions remain completely invisible even at a large system size, which is a striking manifestation of the subtleties that self-organising processes in a complex system may evoke. Before presenting these results in more detail, however, we first proceed with the description of the public goods game with diverse levels of tolerance.

\section{Public goods game with diverse tolerance levels}
The public goods game is staged on a square lattice with periodic boundary conditions where $L^2$ players are arranged into overlapping groups of size $G=5$ such that everyone is connected to its $G-1$ nearest neighbours. Accordingly, each individual belongs to $g=1,\ldots,G$ different groups. The square lattice is the simplest of networks that allows us to go beyond the well-mixed population assumption, and as such it allows us to take into account the fact that the interactions among players that are engaged in group interactions are often inherently structured rather than random \cite{perc_jrsi13}. Nevertheless, for comparison and to provide some primer intuition about the competition of tolerant strategies, we also present results obtained in a well-mixed population, where groups are formed randomly.

Initially each player on site $x$ is designated either as a defector ($s_x = D$), a cooperator ($s_x = C$), a loner ($s_x = L$), or a tolerant player ($s_x = M_i$). Evidently, there are as many levels of tolerance as there are possible defectors in the group, so that $i=0, \ldots, G-1$. If the number of defectors in a group is smaller than $i$ the player $M_i$ acts as a cooperator, while otherwise it acts as a loner. As such, the value of $i$ determines the level of tolerance a particular $M_i$ player has. The higher the value of $i$, the higher the number of defectors that are tolerated by an $M_i$ player within a group without it refusing cooperation. As the two extreme cases, $i=0$ indicates that an $M_i$ player will always remain in the non-participatory loner state, while $i=G-1$ indicates that an $M_i$ player will switch to a loner state only if all the other neighbors are defectors. Importantly, regardless of the choice an $M_i$ player makes, it always bears the cost $\gamma$ as a compensation for knowing the number of defectors in a group.

As is standard practice, all cooperative strategies contribute a fixed amount $c$, here considered $c=1$ without loss of generality, to the public good while defectors and loners contribute nothing. The sum of all contributions in each group is multiplied by the synergy factor $r$ and the resulting public goods are distributed equally amongst all the group members that are not loners. Importantly, the $r>1$ factor is applied only if there are at least two contributions made to the common pool from within the group. Otherwise, a lonely contributor is unable to utilize on the synergistic effect of a group effort, and hence $r=1$ applies. It is also an accepted protocol that loners, who do not participate in the game, obtain a moderate but secure payoff $\sigma$. By following previous related research \cite{hauert_s02,hauert_jtb02}, we use $\sigma=1$ but emphasize that using other values does not change our main findings as long as the rank of payoff values remains unchanged.

As we have argued, the behavior of tolerant players depends sensitively on the number of defectors within a group, which we denote as $n_D$. Diverse tolerance thresholds can be introduced by using different $\delta_i$ prefactors, which are $\delta_i=0$ if $n_D \ge i$ and $\delta_i=1$ if $n_D < i$. Hence, the total number of contributors to the common pool is
\begin{equation}
n_{TC} = n_C + \sum_{i=0}^{G-1} \delta_i n_{M_i}\,,
\end{equation}
where $n_s$ denotes the number of players in the group who follow strategy $s$. By using this notation, the payoff values of the competing strategies obtained from each group $g$ are:
\begin{eqnarray}
\Pi_D\,\,\, &=& \frac{r (n_{TC} \cdot c)}{n_D+n_{TC}}\,,\\
\Pi_C\,\,\, &=& \Pi_D - c\,,\\
\Pi_L\,\,\, &=& \sigma\,,\\
\Pi_{M_{i}} &=& \delta_i \Pi_C + (1-\delta_i) \sigma - \gamma\,.
\label{payoff}
\end{eqnarray}

Monte Carlo simulations of the public goods game are carried out comprising the following elementary steps. A randomly selected player $x$ plays the public goods game with its $G-1$ partners as a member of all the $g=1,\ldots,G$ groups, whereby its overall payoff $\Pi_{s_x}$ is thus the sum of all the payoffs $\Pi_{s_x}$ acquired in each individual group. Next, player $x$ chooses one of its nearest neighbours at random, and the chosen co-player $y$ also acquires its payoff $\Pi_{s_y}$ in the same way. Finally, player $y$ adopts the strategy from player $x$ with a probability given by the Fermi function
\begin{equation}
w(s_x \to s_y)=1/\{1+\exp[(\Pi_{s_y}-\Pi_{s_x}) /K]\}\,\,,
\end{equation}
where $K=0.5$ quantifies the uncertainty by strategy adoptions \cite{szabo_pre98,szabo_pr07,szolnoki_pre09c}, implying that better performing players are readily adopted, although it is not impossible to adopt the strategy of a player performing worse. Each full Monte Carlo step ($MCS$) gives a chance to every player to change its strategy once on average. Further regarding the applied strategy update dynamics, we emphasize that only imitation is allowed but mutation is not considered. If we would apply a permanent mutation rate $\nu$, then all competing strategies would be present in the system. In particular, the smallest fraction of a strategy would be at least proportional to the mutation rate $\nu$. Under such circumstances, we can no longer speak about transitions between different solutions, and this effectively makes the presentation of phase diagrams meaningless. One could measure the stationary frequency of every strategy in the presence of mutation, but this is beyond the scope of the present work.

In case of well-mixed populations, we have used $N=10^7$ players, while the system size for the square lattice topology was varied from $L \times L = 100 \times 100$ to $6000 \times 6000$. As we will discuss in detail later, the appropriately chosen system size is essential to get reliable results that are valid in the large size limit. As expected, the relaxation time depends sensitively on both the system size and the proximity to the phase transition points, which is why it was varied from $10^3$ to $10^6$ $MCS$s. We note that the random initial state may not necessarily yield a relaxation to the most stable solution of the game even at such a large system size ($L=6000$). To verify the stability of different subsystem solutions, we have therefore conducted an appropriate stability analysis that involves the usage of prepared initial states, as illustrated in Fig.~\ref{stability}. In the next section, we present the main results.

\section{Results}

\begin{figure}
\centerline{\epsfig{file=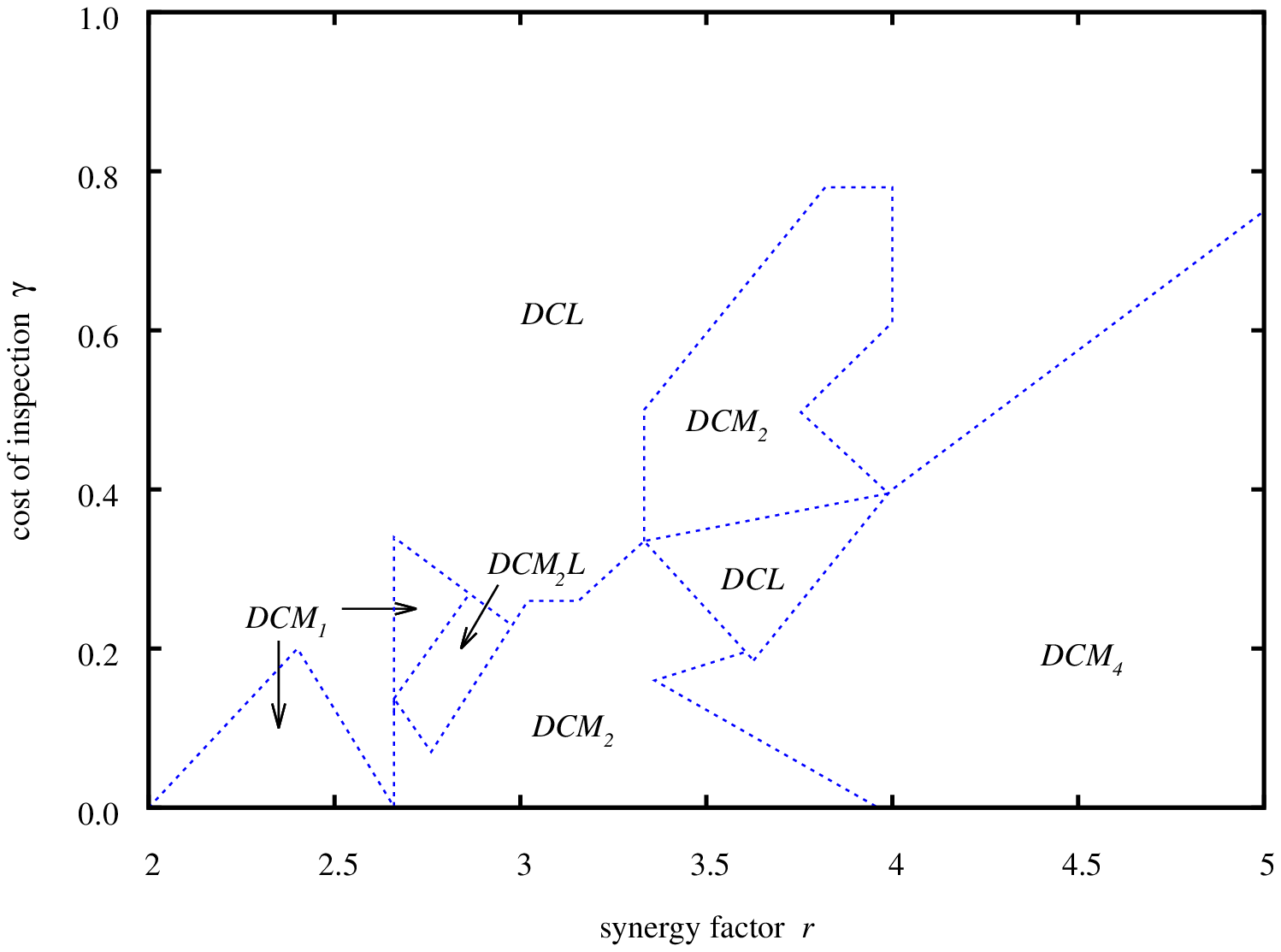,width=12.0cm}}
\centerline{\epsfig{file=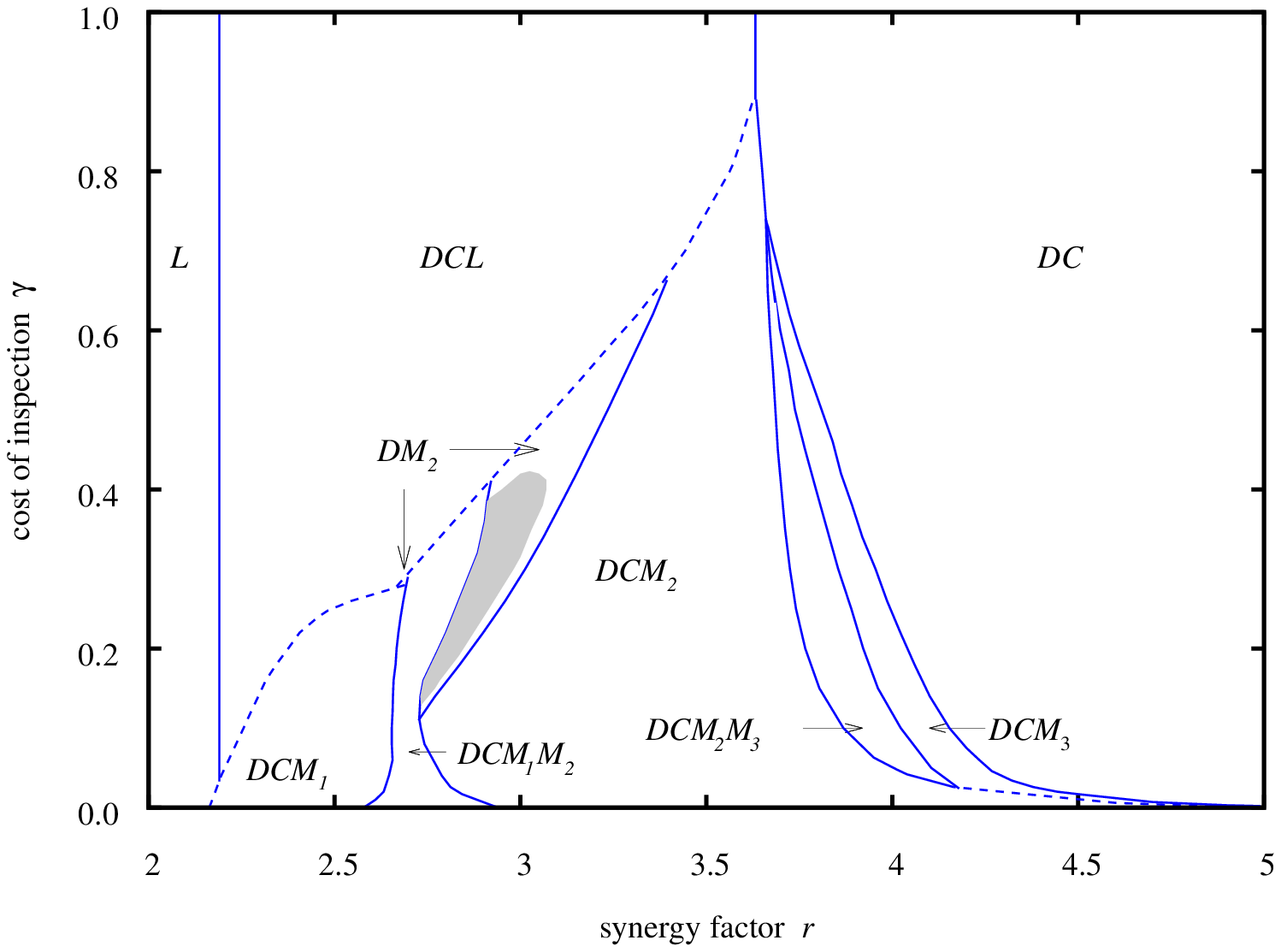,width=12.0cm}}
\caption{\label{phd} Full phase diagram of the $8$-strategy model on the $r-\gamma$ parameter plane, as obtained for a well-mixed (upper panel) and structured population (lower panel).
Dashed lines denote discontinuous, while solid lines mark continuous phase transitions. In the well-mixed population a single tolerant strategy is always selected during the evolution, while in the structured population there also exist regions where the coexistence of different tolerance levels is the stable solution. In the latter case the grey-shaded region in the two-strategy $DM_2$ phase denotes those parameter values where the population evolves into a full-cooperator, defector-free $M_1M_2$ phase, but only if the system size is large enough. We attend to these invisible solutions in detail in Fig.~\ref{size}.}
\end{figure}

Before presenting the main results obtained in structured populations, we first present results obtained in well-mixed populations, where players do not have fixed neighbours but rather form groups randomly. Due to the large number of independent variables analytical solutions are unattainable, but we can perform individual-based simulations to explore different outcomes of the evolutionary process. It is a well-known fact that the behaviour of a well-mixed system may depend sensitively on the initial conditions. We therefore restrict ourself to the case when all competing strategies are initially present in equal proportions. Moreover, to further simplify the calculation, we applied deterministic strategy updating in this case, such that a player adopts the strategy of a randomly chosen partner only if the latter achieved a higher payoff. Despite of these simplifications, the outcome of evolution still remains ambiguous in a well-mixed population. More precisely, there are parameter values where the final outcome is different despite the fact that we have used exactly the same parameters and initial conditions, and only different seeds for the random number generator (the latter provides an inseparable source of noise that influences the outcome of the evolution). While this behaviour is more pronounced in smaller populations, we could still observe it even when using the largest available system size ($N=10^7$ players). Consequently, it is impossible to present a rigorously valid phase diagram that depicts the outcome of the system unambiguously at specific parameter values. Nevertheless, the analysis does provide an intuition about the competition of the strategies at different parameter values. More precisely, the upper panel of Fig.~\ref{phd} depicts the most commonly selected stable solutions, as obtained for a given combination of the synergy factor $r$ and the cost of inspection $\gamma$. This statistically valid diagram was obtained by using $N=10^7$ players and independently repeating the evolution $100$ times for every point of the parameter plane. The presented results indicate that the competition among $M_i$ strategies will always select a sole winner that can coexist with unconditional $C$ and $D$ players. Also, it can be observed that the lower the synergy factor $r$, the lower the threshold from the winning tolerant strategy. These results thus suggest that, under harsh conditions, tolerant players do not have the luxury of tolerating too many defectors in the group because the small value of $r$ simply cannot compensate their lack of contributions to the public good. The reverse is also true, namely that at larger values of $r$ the $M_i$ strategy with a higher threshold outperforms the less tolerant strategies. In the extreme case, when $\gamma$ is also high, then all conditionally tolerant strategies extinct and unconditionally cooperators coexist with defectors due to the well-known consequence of network reciprocity: if $C$ players manage to form compact clusters then their resulting payoff could be high enough to compensate the presence of defectors at the border of such domains. In this way we can observe a pattern where cooperator domains are separated by defector players where their time-independent fractions depends sensitively on the value of $r$.

As was frequently observed for various other evolutionary games before \cite{szabo_pr07,perc_bs10,szolnoki_jrsif14,szabo_pr16}, going from well-mixed to structured populations significantly affects the results, and the presently considered model is in this regard no different. Importantly, the outcome of the evolutionary process in a structured population is always unambiguous if only the system size is large enough. In other words, if a solution that emerges somewhere in the population is capable to dominate other solutions, it will gradually spread and prevail in the whole system. In this way, the original meaning of the phase diagram is restored, as it informs us about the stable solutions that are obtained in dependence on the main system parameters. For the square lattice, such a phase diagram is shown in the bottom panel of Fig.~\ref{phd}. The presented results reveal several fundamental features that can be associated with the viability of tolerant strategies in the public goods game. In agreement with the preceding results obtained in well-mixed populations, it can be observed that the higher the value of $r$, the higher the tolerance can be, and vice versa. This observation resonates with our naive expectation and perception of tolerance in that overly tolerant strategies cannot survive in the presence of other less tolerant strategies. From the viewpoint of the considered evolutionary game this is not surprising, because players adopting the $M_4$ strategy act as loners only if everybody else in the group is a defector. And such sheer unlimited tolerance is simply not competitive with other less tolerant strategies. Also, if the cost of inspection is too high, or if the value of the synergy factor is either very low or very high, then tolerant players cannot survive even if they exhibit different levels of tolerance. This observation is in agreement with preceding research on the subject, where only uniform tolerance levels were considered \cite{szolnoki_pre15}.

More precisely, as the lower panel of Fig.~\ref{phd} reveals, at very low values of $r$, similarly as in the simplest three-strategy $DCL$ model, the loners prevail. At slightly larger values of $r$, this single-strategy phase gives way to the three-strategy phase where $D, C$, and $L$ strategies dominate each other cyclically. Interestingly, unlike in the uniform tolerance public goods game \cite{szolnoki_pre15}, here the $D \to C \to L \to D$ closed loop of dominance is the only way for loners to survive at higher $r$ values \cite{hauert_s02}. In all the other cases the loners die out due to the fact that the diversity of tolerant players is able to provide a more competitive response to the exploitation of defectors. Indeed, if we increase the synergy factor further, we find that, through a succession of different phase transitions, the $DCL$ phase gives way to a rich variety of two-, three- or even four-strategy phases, in all of which tolerant strategies are present. These solutions are the $DCM_1$, the $DM_2$, the $DCM_2$, and the $DCM_3$ phase, as well as two four-strategy phases $DCM_1M_2$ and $DCM_2M_3$, which are unique to the public goods game with diverse strategy levels. We emphasize that the coexistence of tolerant strategies with different tolerance levels is due entirely to the consideration of structured populations. We remind that such evolutionary outcomes are impossible in well-mixed populations, as evidence by the results presented in the upper panel of Fig.~\ref{phd}.

\begin{figure}
\centerline{\epsfig{file=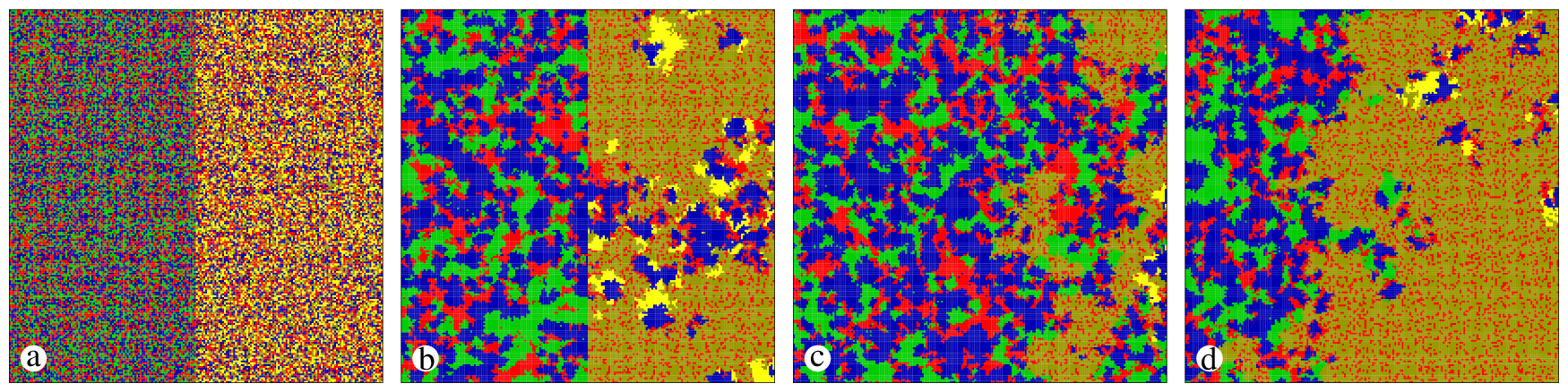,width=17cm}}
\centerline{\epsfig{file=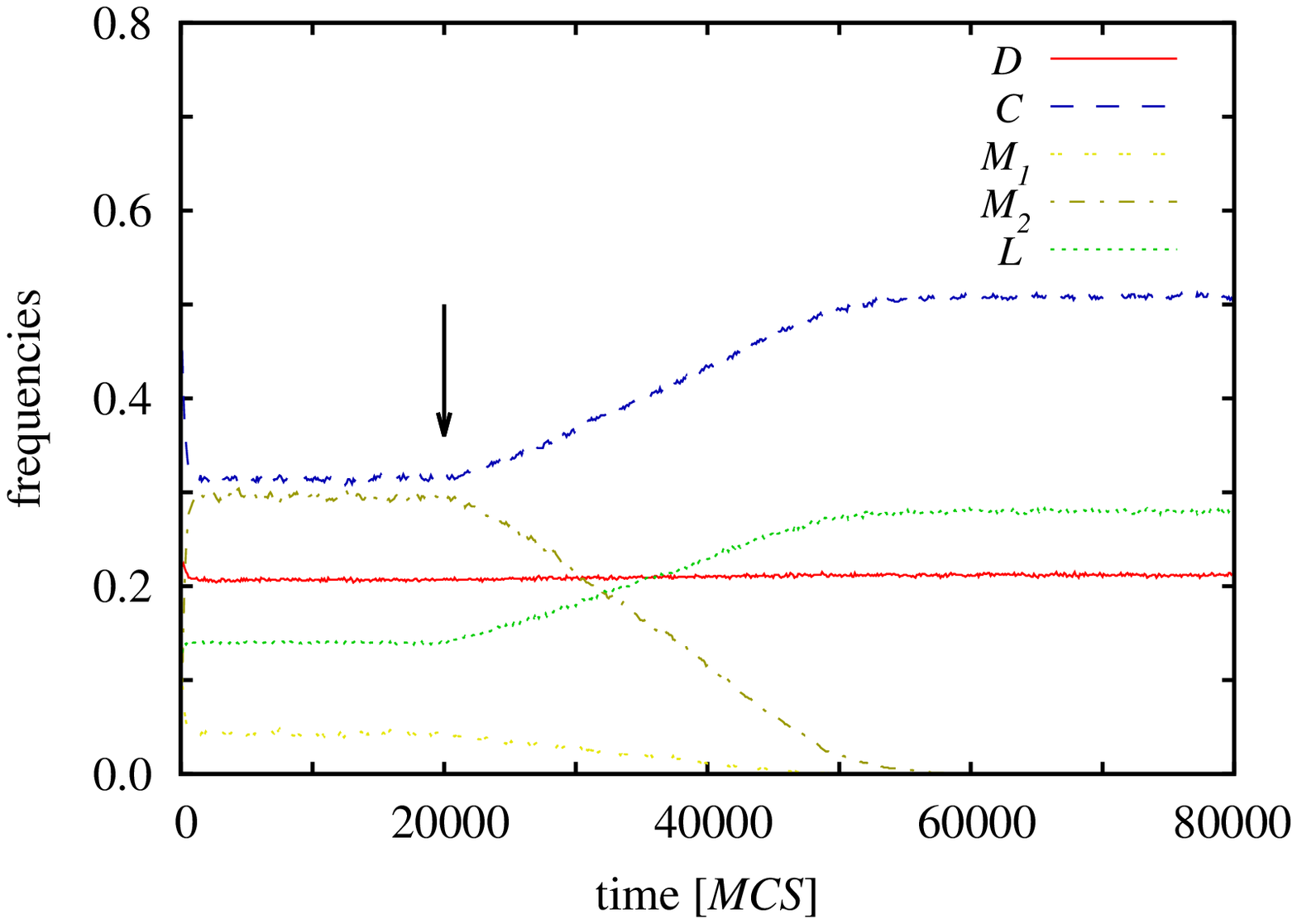,width=8cm}\epsfig{file=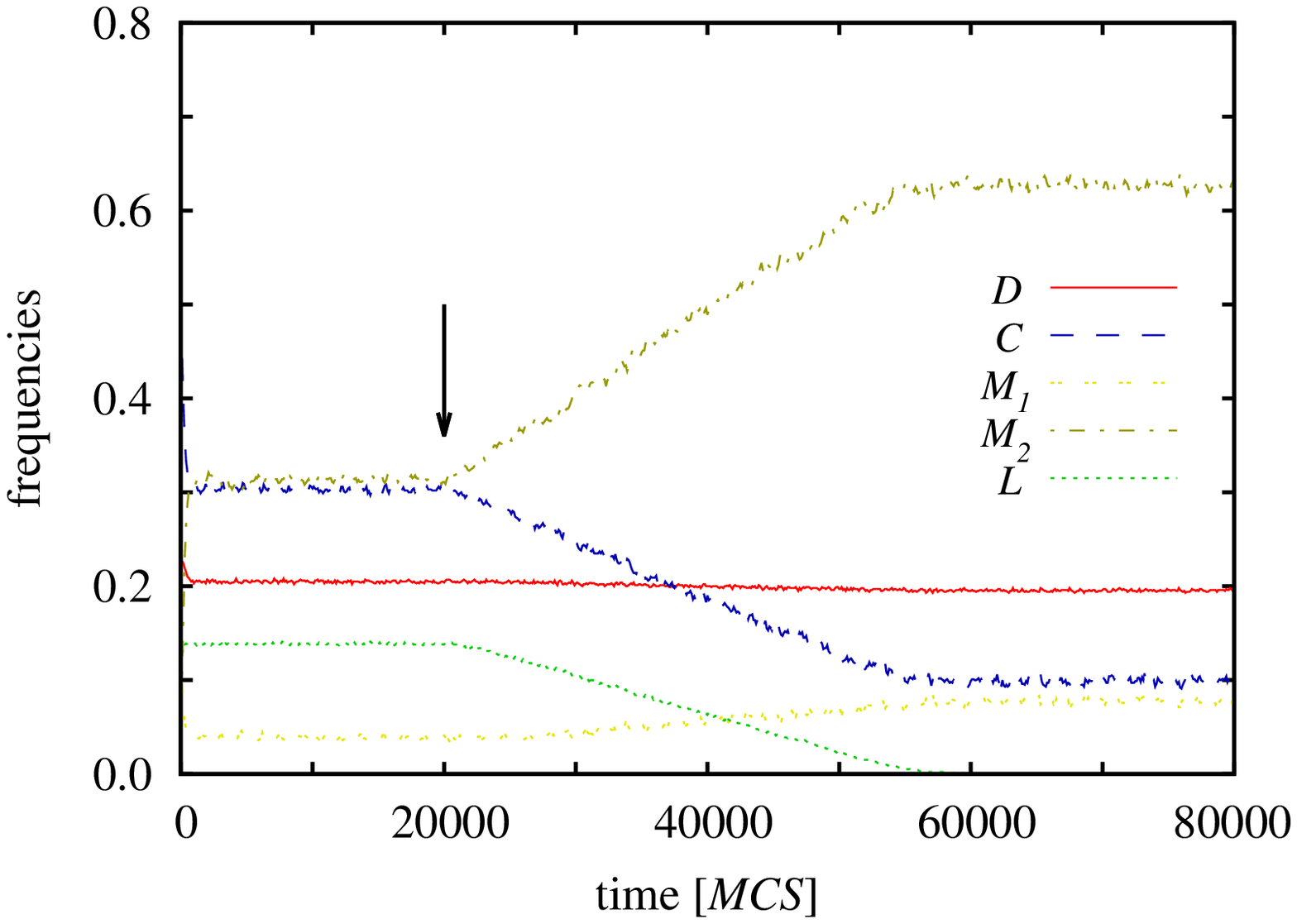,width=8cm}}
\caption{\label{stability} An example of a proper stably analysis of two subsystem solutions, namely the three-strategy $DCL$ and the four-strategy $DCM_1M_2$ phase, which are separated by a discontinuous phase transition in Fig.~\ref{phd}. The series of snapshots in the upper row shows the separation of the lattice in two parts, each of which is initially randomly populated with the strategies that will form one of the two competing subsystem solutions [panel~(a)]. In panel (b), the subsystem solutions are formed in both halves of the square lattice, and accordingly, their competition can start by removing the border between them (thus allowing strategy transfer across the border). Panel (c) shows an intermediate state during the competition in which the $DCL$ phase will ultimately turn out to be the winner. Here $r=2.80$ and $\gamma=0.35$. Panel (d), on the other hand, shows an intermediate state during the competition in which the $DCM_1M_2$ phase will ultimately turn out to be the winner. Here $r=2.81$ while the value of $\gamma$ is unchanged. Importantly, for both values of $r$ the state depicted in panel (b) is qualitatively exactly the same (both phases are individually stable regardless of which value of $r$ is used). The two graphs in the bottom row depict the corresponding ($r=2.80$ left and $r=2.81$ right) time evolution of the strategy densities. After a relaxation of 20~000 $MCS$ (marked by an arrow), the two subsystem solutions start competing for space. On the left side the $DCL$ solution wins, while on the right side the $DCM_1M_2$ solution wins. The linear system size used for this example was $L=2400$, but the snapshots in the upper row contain just a $200 \times 200$ cutoff of the whole population for clarity.}
\end{figure}

According to the phase diagram in the bottom panel of Fig.~\ref{phd}, the first transition from the $DCL$ phase to one of the mentioned phases is always discontinuous. The accurate position of these kind of phase transition points, as denoted by the dashed line in the phase diagram, can only be determined by means of a stability analysis of competing subsystem solutions. A subsystem solution can be formed by any subset of all the competing strategies, and on their own (if separated from other strategies) these subsystems solutions are stable. This is trivially true if the subsystem solution is formed by a single strategy, but is likewise true if more than one strategy forms such a solution. Evidently then, for any specific set of parameter values, more than one subsystem solution exists. The dominant subsystem solution, and hence the phase that is ultimately depicted in the phase diagram such as the one shown in the bottom panel of Fig.~\ref{phd}, can only be determined by letting all the subsystem solutions compete against each other. As an example, we show in Fig.~\ref{stability} the competition between the three-strategy $DCL$ phase and the four-strategy $DCM_1M_2$ phase at both sides of the discontinuous phase transition point, although on both sides the two phases are individually stable, i.e., are proper subsystem solutions. To monitor the competition between the two subsystem solutions, we launch the evolution from a prepared initial state, where one half of the lattice initially contains only strategies $D, C$, and $L$ distributed uniformly at random, while the other half of the lattice contains only the strategies $D, C, M_1$, and $M_2$ distributed uniformly at random. As the next step, we let the two subsystem solutions evolve to their representative state in terms of the strategy frequencies and the typical spatial pattern. Only when both reach their stationary state we open the border by allowing strategy invasion through the separating interface. Lastly, we monitor how the competition between the two solutions evolves, i.e., which subsystem solution will turn out as the victor. The example depicted in Fig.~\ref{stability} demonstrates clearly that the final outcome depends sensitively on the value of the synergy factor $r$. At the smaller value of $r$ the $DCL$ solution wins, while at the slightly larger value of $r$ the $DCM_1M_2$ solution turns out to be the dominant one.

We emphasize that finite-size effects can easily play an obstructive role in the processes illustrated in Fig.~\ref{stability}. If we start the evolution from a random initial state using a small system size, it can easily happen that we observe a misleading evolutionary outcome, simply because the phase that would be a stable solution at large system size has no chance to emerge -- for example, one of the strategies that would be necessary to form it dies out beforehand due to the small system size. But that is not the only caveat. Even if we use prepared initial states for the stability analysis, we should be careful because the space (part of the lattice) allocated to each potential subsystem solution should be large enough for the latter to emerge. For example, the fluctuations of strategies in the cyclically dominant $DCL$ phase could be extremely large, and therefore this subsystem solution alone requires a large population to avoid fixation before the characteristic stationary pattern emerges. We note that the upper panels of Fig.~\ref{stability} show just a small patch around the border where the two solutions meet, which is cut out of a large $2400 \times 2400$ lattice (not shown). This is also why the periodic boundary conditions cannot be detected in the four depicted snapshots.

\begin{figure}
\centerline{\epsfig{file=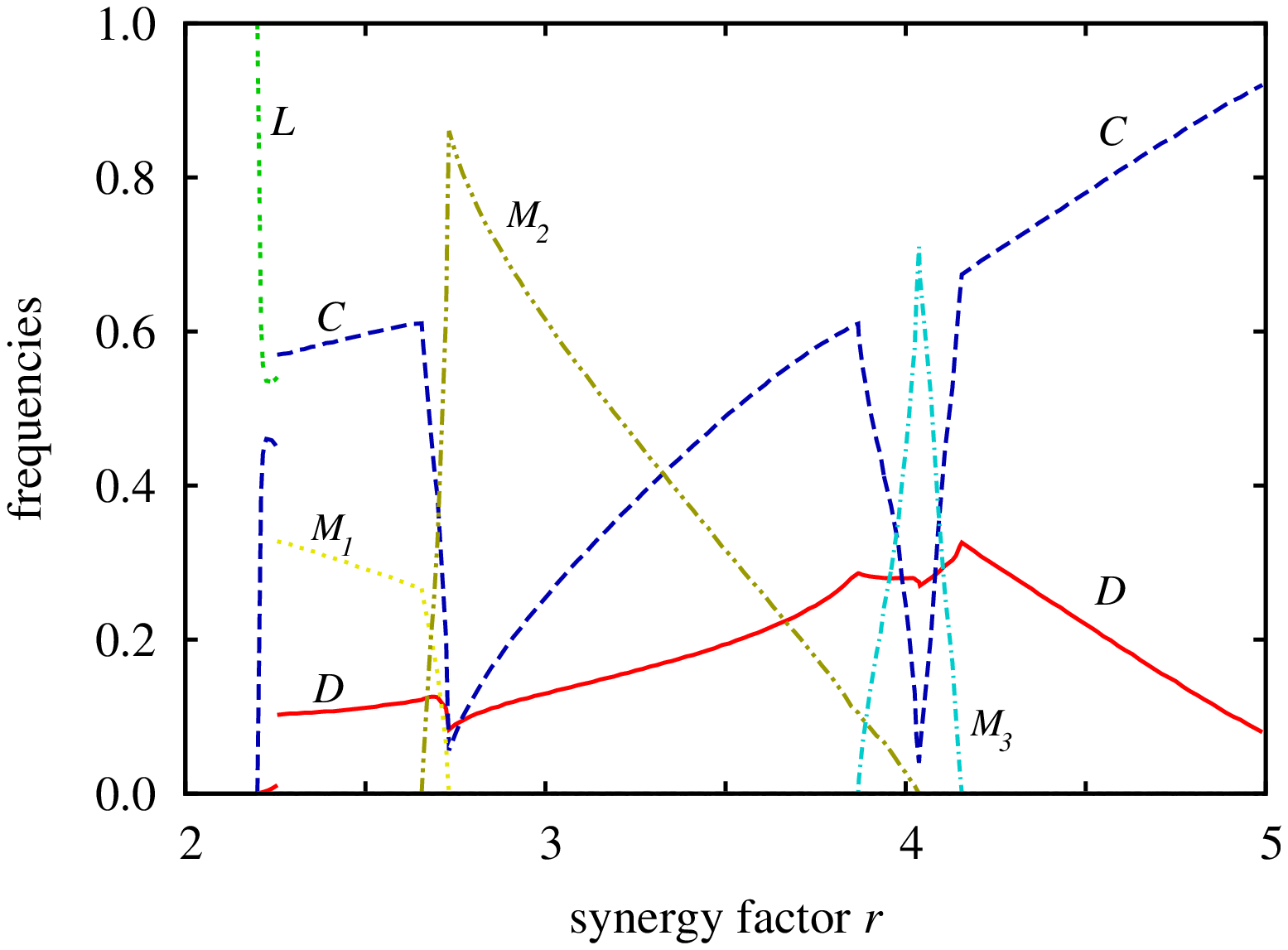,width=8cm}\epsfig{file=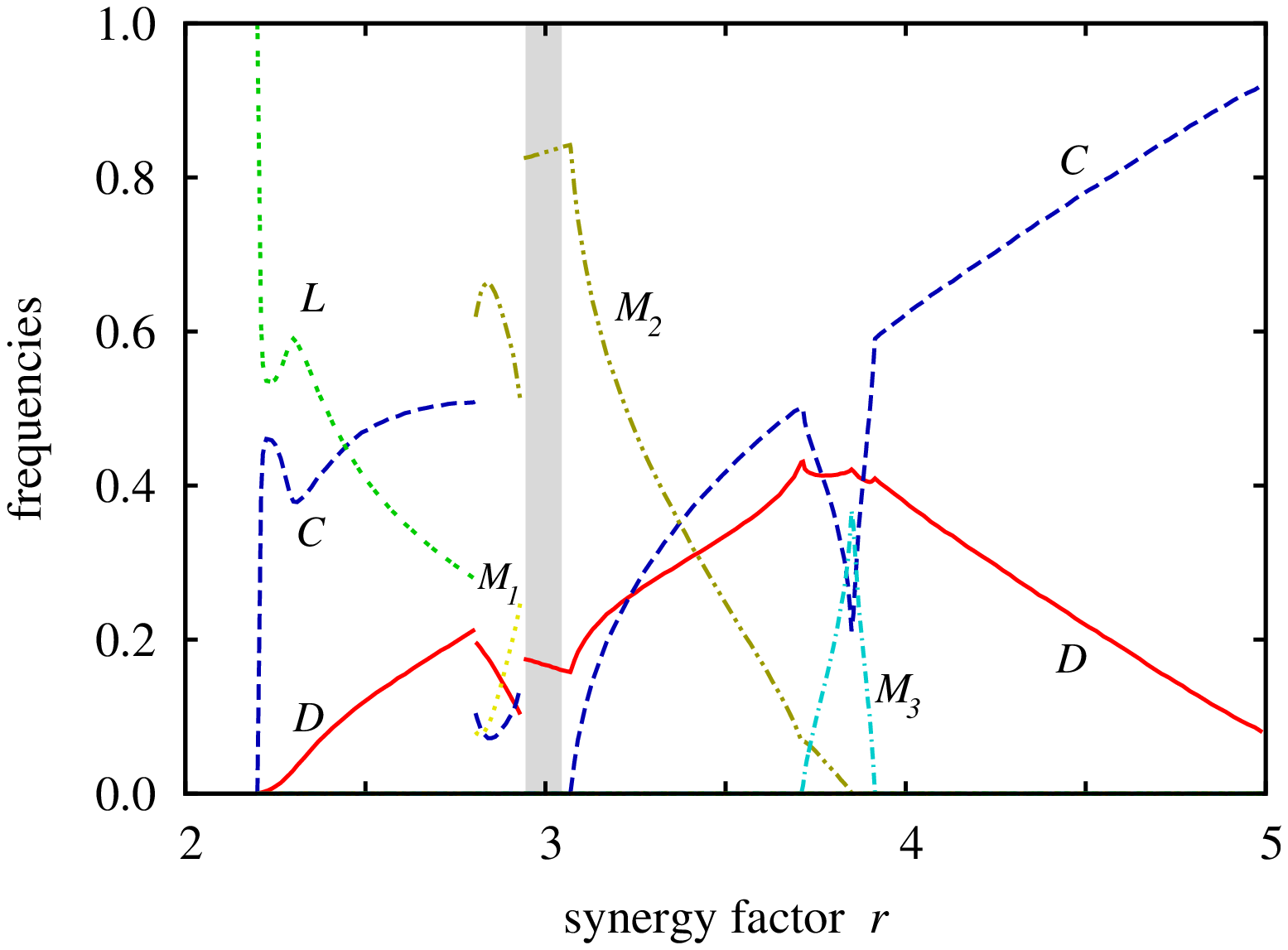,width=8cm}}
\caption{\label{cross} Two representative cross-sections of the phase diagram depicted in Fig.~\ref{phd}, left as obtained for $\gamma=0.10$ and right as obtained for $\gamma=0.35$. In both cases there exist values of $r$ at which different tolerant strategies can coexist, and in both cases the population goes through a succession of continuous and discontinuous phase transitions. For $\gamma=0.35$ the four-strategy $DCM_1M_2$ phase becomes unstable as we increase $r$. The next phase is the two-strategy $DM_2$ phase. Importantly, however, this phase can only be reached if the system size is relatively small (see Fig.~\ref{size} for details). If the system size is sufficiently large, then the evolution terminates into a full-cooperator, defector-free $M_1M_2$ phase, which is marked by the grey-shaded interval. The error bars in both plots are comparable to the width of the lines.}
\end{figure}

We proceed with the exploration of the main phase diagram obtained on the square lattice, and to have a more accurate insight with regards to the recorded solutions, we show in Fig.~\ref{cross} two representative cross sections. The presented results reveal that there is a clear selection among the different tolerant strategies, and hence often only one of them survives and forms a stable solution with either the $D$ or~/~and the $C$ players. There are, however, also specific parameter values where the stable solution contains two different tolerant strategies. While an explanation for such a coexistence will be presented later, the message is that, sometimes, diverse levels of tolerance do provide an optimal response to the public goods dilemma.

A more accurate comparison of the two plots in Fig.~\ref{cross} reveals qualitatively different behaviour. When the cost $\gamma$ is low (left), we can observe continuous transitions between the phases. Starting from the $DCM_1$ phase, we first see that the fraction of the $M_2$ players increases gradually while the fraction of the $M_1$ players decreases. Similarly, when we leave the $DCM_2$ phase, the fraction of the $M_3$ players in the population grows while the fraction of the $M_2$ players decreases. Finally, when the synergy factor provides a sufficiently comfortable environment for pure cooperators, the last tolerant strategy ($M_3$) dies out, giving way to the $DC$ phase. Intuitively, one might expect that a higher inspection cost would make the tolerant strategies more vulnerable. Yet this is true only for very high $\gamma$ values. The left panel of Fig.~\ref{cross} illustrates that the four-strategy $DCM_1M_2$ phase terminates qualitatively differently from the way it does for the low $\gamma$ value. In the high-$\gamma$ case the time average of the fraction of each strategies is well above zero, but the four-strategy phase becomes unstable as we increase $r$ due to invincible fluctuations of strategy frequencies.

\begin{figure}
\centerline{\epsfig{file=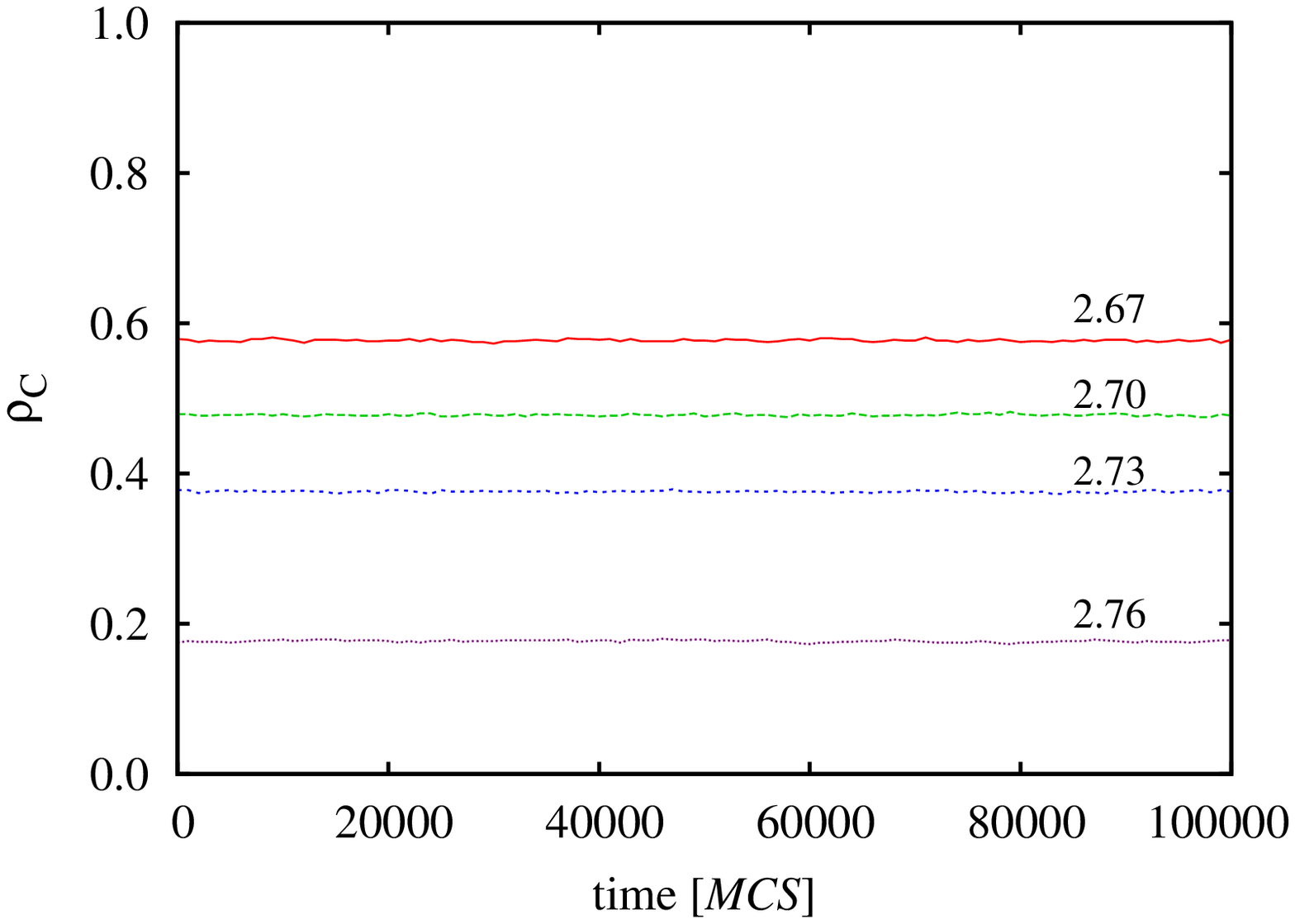,width=8cm}\epsfig{file=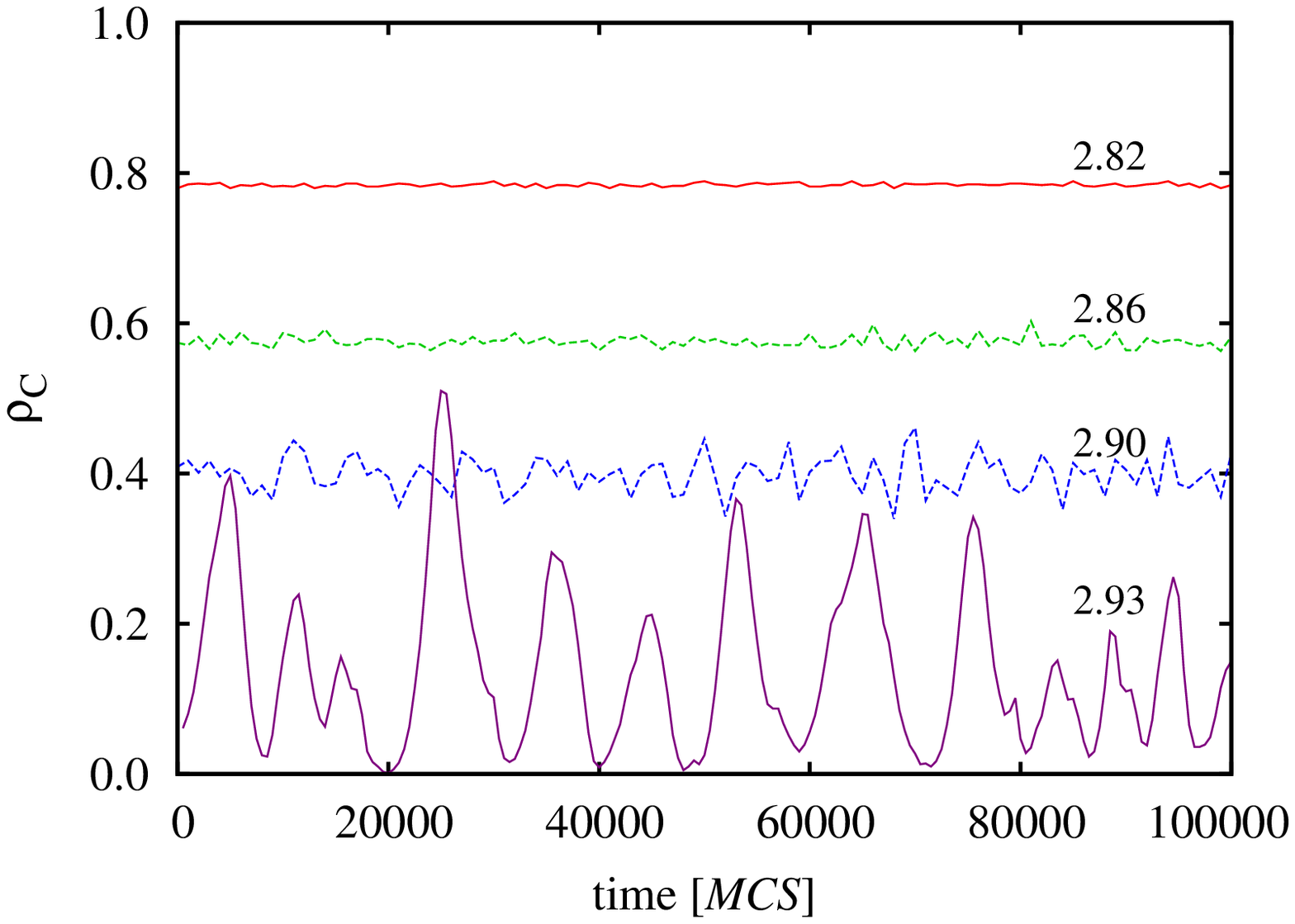,width=8cm}}
\caption{\label{fluct} The time course of the fraction of cooperators in the stationary state, as obtained for different values of the synergy factor $r$ (see legend) at $\gamma=0.05$ (left) and at $\gamma=0.35$ (right). For all the cases we have used a large $4800 \times 4800$ lattice, which ought to result in smooth lines in all the cases, as indeed it does in the left panel, and for some values of $r$ in the right panel. However, as we approach the phase transition point with larger values of $r$, it can be observed that the amplitude of the fluctuations in the right panel increases significantly. Most interestingly, these fluctuations are not because the average fraction of cooperators would go to zero as we increase $r$. In fact, the right panel of Fig.~\ref{cross} illustrates the opposite, which is that $\rho_C$ slightly increases as we approach the critical point. The comparison thus indicates that the four-strategy $DCM_1M_2$ phase becomes inherently unstable as we increase $r$ at $\gamma=0.35$. For clarity, the upper three curves in the right panel are shifted vertically by a constant value.}
\end{figure}

This behaviour is illustrated in Fig.~\ref{fluct}, where we show how the fraction of cooperators varies over time for different values of $r$ at two different values of $\gamma$. To emphasise the remarkably different behaviour, we have used a notably large system size, $L=4800$, where normally the fluctuations of strategy densities should be virtually invisible. This expectation is confirmed in the left panel when the system leaves the four-strategy $DCM_1M_2$ phase via a continuous phase transition. But in the right panel, when a higher cost $\gamma$ is applied, the fluctuations in amplitude are increasing drastically as we increase the value of $r$ and approach the discontinuous phase transition point.

In this ``unstable'' parameter region the final solution depends sensitively on which strategy dies out first. According to our observations, strategies $M_0$, $M_3$, $M_4$, and $L$ die out very soon, and the fundamental question remains whether the $M_1$ strategy or the $C$ strategy will die out next, because this ultimately dictates two very different solutions. If the $M_1$ strategy dies out first, we return to the previously studied uniform-tolerance public goods game \cite{szolnoki_pre15}, where the system terminates into the two-strategy $DM_2$ phase. Indeed, this scenario materialises almost always if the system size is small. If the system size is large, however, the $C$ strategy dies out first, which allows the two tolerant strategies to form the stable two-strategy $M_1M_2$ phase. In this case the presence of diverse tolerance levels thus ensures that defectors are wiped out completely, leaving the population in defector-free, full-cooperator state.

\begin{figure}
\centerline{\epsfig{file=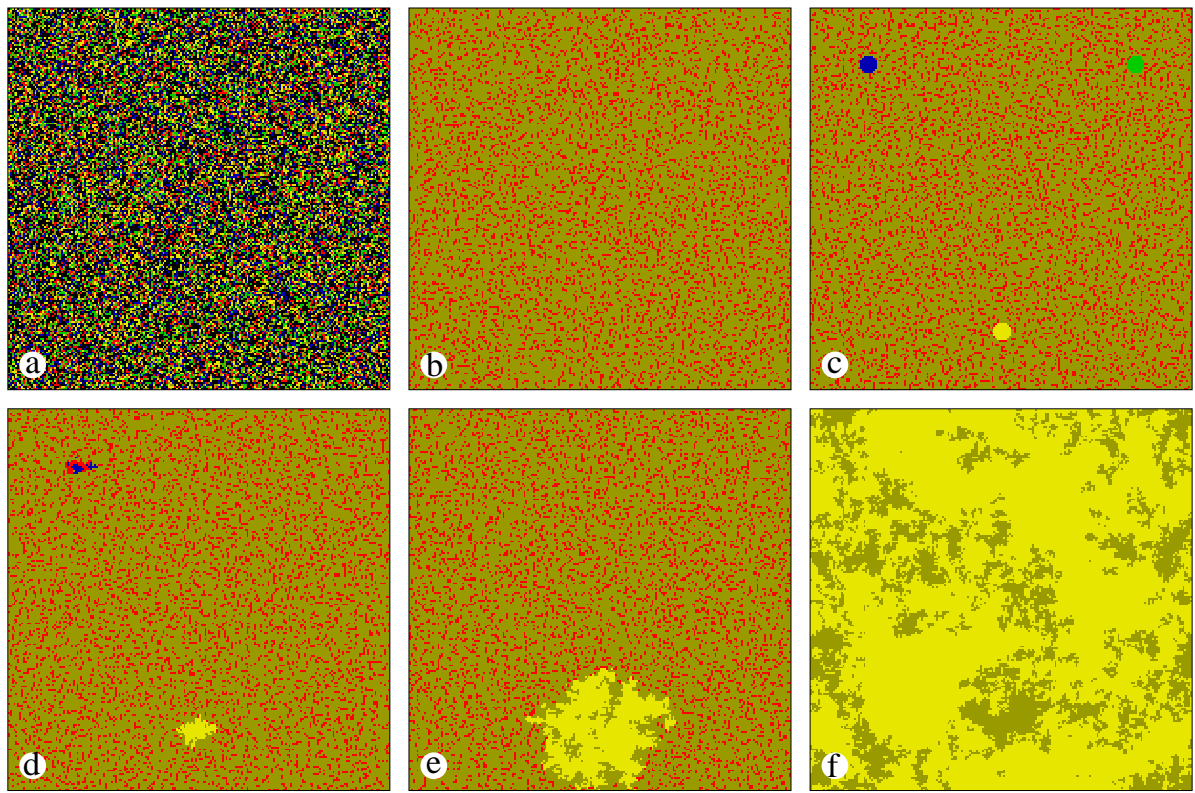,width=16cm}}
\caption{\label{DM2} Finite-size effects may lead to misleading outcomes of the strategy competition. At the beginning, we start the evolution from a random initial state where all the eight strategies are present [panel~(a)]. After the relaxation of $1000$ MCS [panel~(b)], only $D$ (red) and $M_2$ (ochre) strategies survive, which also form the stable two-strategy phase in the uniform-tolerance public goods game at the same parameter values ($r=2.97$ and $\gamma=0.35$). In panel~(c), we then manually re-introduce small compact patches of three other strategies, namely strategy $C$ (blue), strategy $L$ loner (green), and strategy $M_1$ (yellow). Soon afterwards, loners die out (again) very soon within the $DM_2$ phase, as shown in panel~(d). Later, $C$ players also die out, but tolerant strategies $M_1$ and $M_2$ form a successful alliance, as shown in panel~(e). Indeed, the $M_1M_2$ subsystem solution turns out to be stronger than the previously declared victorious $DM_2$ phase, and the population terminates in a defector-free state, as shown in panel~(f). The system size used in this example is $L=200$, while stages presented in panels~(d), (e), and (f) were obtained after $30$, $250$, and $1300$ $MCS$s from panel~(c) onwards. Importantly, as we demonstrate in Fig.~\ref{size}, the defector-free state can evolve naturally from a random initial state if the system size is large enough.}
\end{figure}

To illustrate how the combination of different tolerant strategies is the most effective response to a public goods dilemma, we present a series of snapshots that is carefully engineered to illustrate exactly what otherwise remains invisible if a small population starts from a random initial state. In the panel~(a) of Fig.~\ref{DM2}, we start exactly like that -- with a small $200 \times 200$ lattice where initially all eight strategies are distributed uniformly at random. After a relatively short relaxation time only $D$ and $M_2$ players survive in panel~(b) to form what appears to be the dominating two-strategy phase at these parameter values. In fact, this $DM_2$ phase is the dominant stable solution in the previously studied uniform-tolerance public goods game \cite{szolnoki_pre15}. But in the game studied here, we have other options because of the diverse tolerance levels that we consider. Therefore, to test the stability of the $DM_2$ phase properly, we insert small compact patches of other strategies in panel~(c). These are the pure cooperators ($C$), the loners ($L$), and the tolerant strategy ($M_1$), positioned from the left-upper part of the lattice in a clockwise manner. Subsequently, panels~(d) and (e) demonstrate clearly that first the loners die out, followed by the pure cooperators. However, the $M_1$ strategy forms a powerful alliance with the $M_2$ strategy, whose domain is able to grow, as shown in panel~(e). Gradually, all the defectors are crowded out, and in the final state illustrated in panel~(f), only the two different tolerant strategies remain to provide the maximal cooperation level in the defector-free state. Thus, it turns out that the $M_1M_2$ subsystem solution is ultimately stronger than the $DM_2$ subsystem solution. This outcome can be interpreted so that $M_2$ players are efficient at sweeping out $C$ and $L$ players, who, as the two extreme limits of tolerance, prevent other $M$ strategies to function efficiently. At the same time, $M_1$ players are capable to beat strategy $D$. Accordingly, we need the features of both $M_1$ and $M_2$ players to reach the happy end.

\begin{figure}
\centerline{\epsfig{file=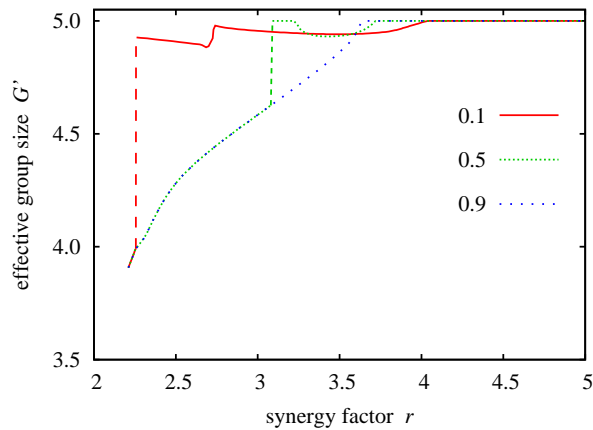,width=8cm}}
\caption{\label{effective_size} The effective size of groups $G'$ when the public goods game is played in dependence on the synergy factor $r$ , as obtained for different values of the cost $\gamma$ (see legend). The presented results demonstrate that the value of $G'$ always exceeds the actual value of the synergy factor $r$, thus always preserving the original public goods dilemma even though the participation is not obligatory. The curve obtained for $\gamma=0.5$ reaches $G'=5$ at an intermediate value of $r$, thereby indirectly revealing why the $M_2$ strategy is particularly effective in the presence of defectors. Namely, players adopting the $M_2$ strategy contribute to the common pool almost exclusively when akin $M_2$ players are also in the group, while otherwise they refuse participation and act as loners.}
\end{figure}

The optional participation in the public goods game can easily result in a smaller effective group size because not only $L$ but also $M_i$ players may refuse participation in the joint venture. This raises the possibility that the fixed value of the synergy factor $r$ becomes larger than the effective group size $G'$, which would transform the public goods dilemma to a game where players would actually have an incentive to prefer cooperation over defection. To clarify this, we have calculated the average size of every group where the public goods game was played, along with the value of the synergy factor $r$ that was used to determine the payoffs. The obtained results are presented in Fig.~\ref{effective_size}, where we show the average value of $G'$ in dependence on the synergy factor $r$, as obtained for three representative values of the cost. Here $\gamma=0.1$, $0.5$ and $0.9$ represent low, intermediate and high cost values of tolerant players, respectively. As the results show, $G'$ is always well-above the corresponding value of $r$, thus leaving the public goods dilemma intact, even if effective group size $G'$ changes. This quantity, however, helps us to gain a deeper understanding of why the above-discussed $DM_2$ coexistence can provide a competitive solution against $C$, $L$ or other $M_i$ strategies. As the middle curve in Fig.~\ref{effective_size} shows, the value of $G'$ jumps up and approaches the maximal $G=5$ level at an intermediate value of $r$, where we have observed the stable $DM_2$ phase. At this point we note again that $G'$ is measured only for those groups where the public goods game is actually played. Accordingly, the high value of $G'$ indicates that $M_2$ players contribute to the common pool almost exclusively when players of the same type are present dominantly in the group. In this case they can collect the highest mutual payoff, otherwise, when they form groups with $D$ players, they refuse to participate and prefer a safer $\sigma$ payoff that is warranted by the loner strategy. In the latter case the public goods game does not take place, and hence the average value of $G'$ is not lowered. Although it is possible to claim that other $M_i$ strategies behave similarly, the intermediate tolerance threshold level applied by the $M_2$ strategy is the best compromise under this conditions to reach the most competitive payoff.

\begin{figure}
\centerline{\epsfig{file=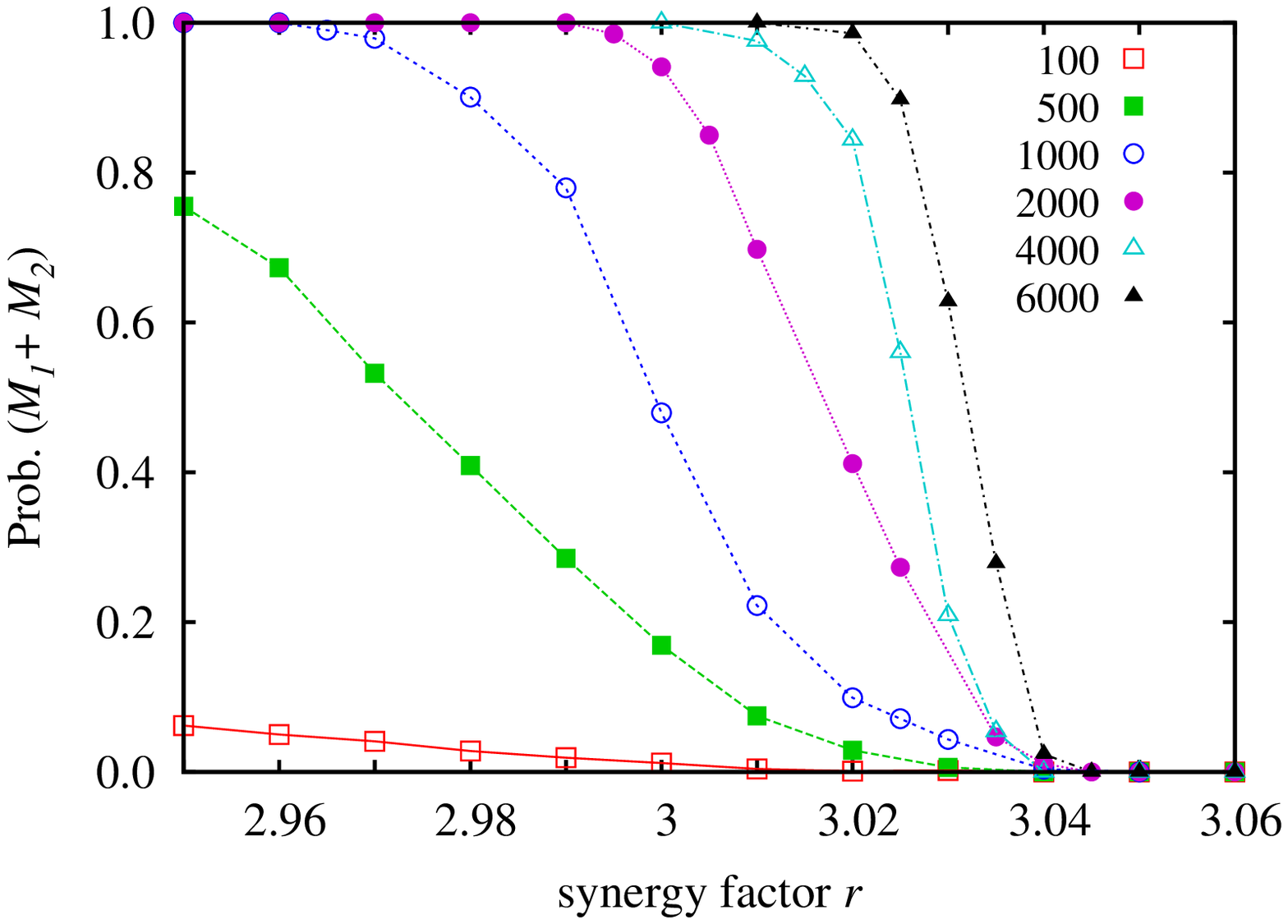,width=8cm}\epsfig{file=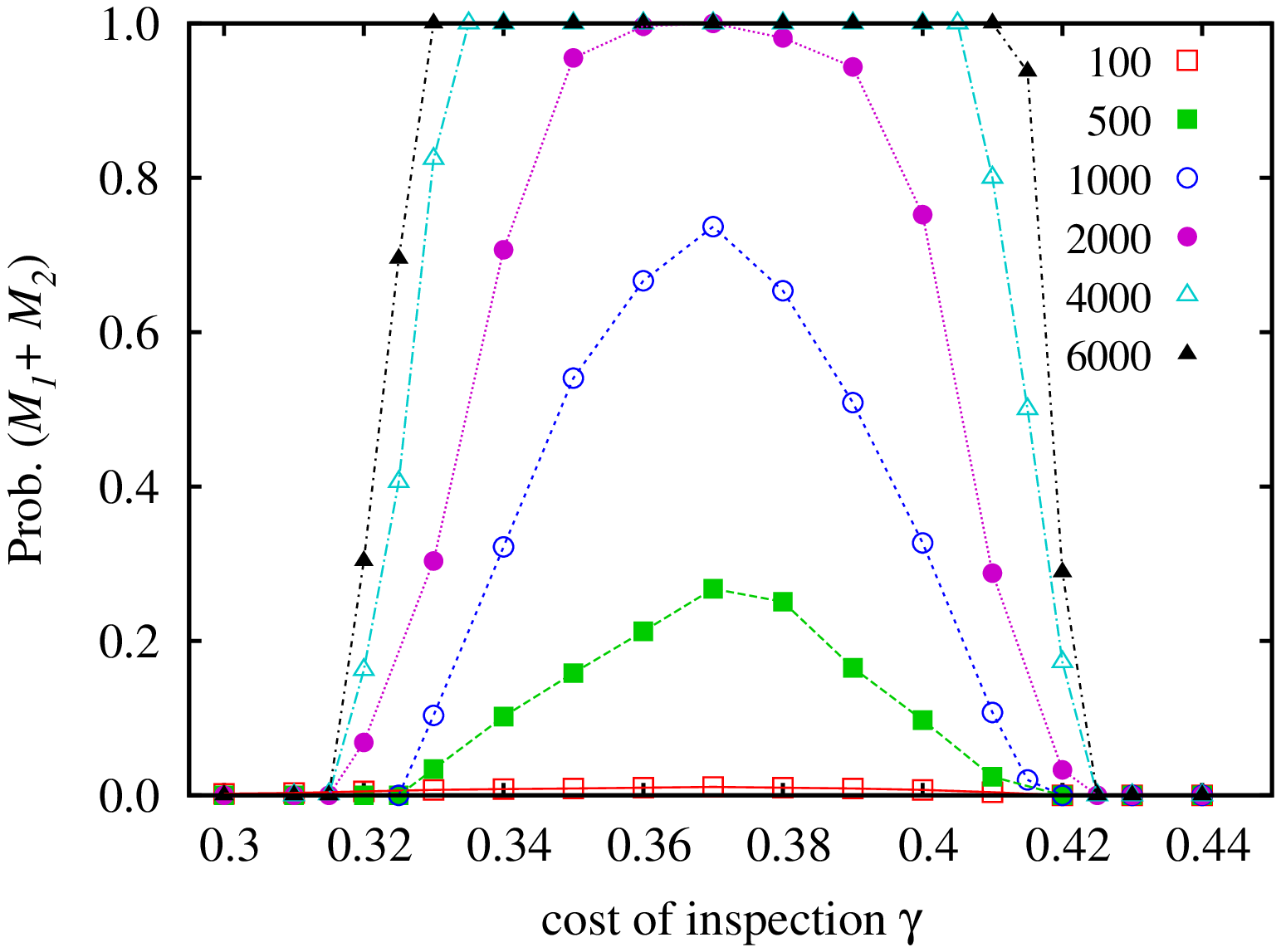,width=8cm}}
\caption{\label{size} The invisible solutions. Fixation probability for the $M_1M_2$ phase, as obtained for $\gamma=0.35$ (left) and for $r=3.0$ (right) in dependence on the other key parameter, respectively. The linear size of the applied square lattices are indicated in the legend. Remarkably, there exist combinations of parameters $r$ and $\gamma$ at which, even at $500 \times 500$ system size, the $M_1M_2$ solutions remains completely invisible, even though at a much larger $6000 \times 6000$ system size the fixation probability for the exact same solution is 1. Depicted probabilities are averaged over 500-1000 independent runs. Lines are just to guide the eye.}
\end{figure}

Turning back to the properties of the phase diagram depicted in the bottom panel of Fig.~\ref{phd}, the example shown in Fig.~\ref{DM2} highlights that the order in which the strategies die out is crucial for the final evolutionary outcome. To give a more quantitatively accurate prediction of whether the system will terminate into the $DM_2$ or the $M_1M_2$ phase, we show in Fig.~\ref{size} the fixation probability to the latter for different combinations of $r$ and $\gamma$. Here, we have always started the evolution from a random initial state (containing all the eight strategies distributed uniformly at random), and we have simply counted how many times the system terminates into the $M_1M_2$ phase. The alternative destination of the evolutionary process was to arrive to the $DM_2$ phase, which for all the considered parameter combinations used in Fig.~\ref{size} is the only other stable subsystem solution. To get a reliable statistics, we have repeated these runs $1000$ times for $L=100-1000$ linear system sizes, while for even larger lattices we have averaged over 500 independent runs. The results depicted in Fig.~\ref{size} show a remarkable finite-size effect. More precisely, there are parameter values of $r$ and $\gamma$ where the strategy $C$ dies out first, thus allowing the $M_1M_2$ phase to conquer the whole system. But this solution remains completely invisible if the system size is too small. And these are the specific combinations of the parameters $r$ and $\gamma$ that are shaded grey in the phase diagram shown in Fig.~\ref{phd} -- these are the invisible solutions for the large majority of the system sizes that are in widespread use in the literature nowadays.

\section{Discussion}
Summarising, we have studied the evolution of cooperation in a spatial public goods game entailing traditional cooperators, defectors, and loners, as well as five different types of tolerant players. Our research has confirmed the naive expectation that minimal tolerance levels are best applied under adverse conditions for cooperation, while higher tolerance levels are preferred under conditions favourable to cooperation. More importantly, we have shown that the diversity of tolerance can give rise to synergistic effects, such that only players with a different threshold provide the optimal response to defection and default abstinence, and in fact become dominant where otherwise, in the absence of tolerance, cooperation would not be viable. The results also reveal, however, that more commonly the evolution selects a single tolerance level to coexist with one or more of the three traditional strategies. This highlights the subtle balance between punishment and reward that must be met for tolerance to have the desired effect. Namely, when a tolerant player decides to abstain, it effectively punishes cooperators by not cooperating with them, and it also punishes defectors by denying them the possibility to exploit a cooperative act. On the other hand, when a tolerant player decides to cooperate, if effectively rewards the defectors by giving them an opportunity to exploit, and it also rewards other cooperators by supporting the joint enterprise. The right balance between the two choices, it turns out, is subtle and difficult to pinpoint, and as such it favours a single tolerance level over a multitude of them. Yet, as we have shown, this does not preclude diverse tolerance levels to constitute an optimal response to the public goods dilemma.

In addition to the more general observations related to the viability of tolerance and its evolutionary advantages, the reported results reveal one of the most striking system-size effects. Namely, we have shown that some globally stable subsystem solutions, and in particularly those that manifest synergies between different tolerance levels, can remain completely invisible even at a relatively large system size. It is only when a sufficiently large system size is used do these invisible solutions sometimes emerge as the victors of the evolutionary process. And only if the system size is really very large does to fixation probability for these invisible solution tends to one. This carries an important warning that is frequently overlooked when random initial conditions are used in structured populations where competing strategies are three or more. Although the usage of random initial conditions may appear as the most 'democratic' setup, giving every strategy an equal opportunity to succeed, in reality it is not the strategies that compete, but rather stable subsystem solutions. Subsystem solutions are solutions that are formed by different subsets of all the competing strategies. In addition to which strategies form a particular subsystem solution, each such solution is characterised by a particular spatiotemporal dynamics that supports its stability. Most importantly, some subsystem solutions therefore require longer than others to emerge, or they require a larger system size than other, usually simpler subsystem solutions. The more complex subsystem solutions therefore often never even get a chance to compete against the faster emerging simpler subsystem solutions. It is thus absolutely vital that the stability of subsystem solutions is verified for all possible strategy combinations, and that the competition among the stable solutions is carried out so that all get a fair chance, as demonstrated in Fig. 3.

It is also important to note that the argument claiming that such demanding subsystem solutions are not relevant because they simply require an unrealistically large system size is wrong. Once such solutions emerge, they can remain stable also in very small populations. The large system size is initially needed simply to give a chance to all possible subsystem solutions to fully emerge, and then for all of them to engage in a fair evolutionary competition. In effect, the usage of a sufficiently large system size simply ensures that all imaginable initial conditions will be realised before an accidental extinction, even if the latter is to occur only locally, would prevent the emergence of a stable subsystem solution. But in reality, the realisation of a particular initial condition, however special it may be, is mainly a matter of time, and in case of intelligence also a matter of craftsmanship towards initiating a particular change, rather than a matter of vast population size. Arguments against  solutions that are valid and remain stable in the large population size limit are futile, as they go against the fundamental principles that have been established a long time ago in the realm of nonequilibrium statistical physics and the theory of phase transitions \cite{liggett_85,binder_88}. And it is of paramount importance for the research relying on simulations of complex systems in biology, ecology, sociology, and economy to adhere to these fundamental principles if it aspires to have any merit.

\ack
We thank our referees for constructive comments. This research was supported by the Hungarian National Research Fund (Grant K-120785) and the Slovenian Research Agency (Grants P5-0027 and J1-7009).

\section*{References}

\providecommand{\newblock}{}

\end{document}